\documentclass[aps,prb,preprint,groupedaddress]{revtex4}
\usepackage[dvipdfmx]{graphicx}
\usepackage{color}
\begin{document}


\title{Magnetism and superconductivity in Sr$_2$VFeAsO$_3$  revealed by $^{75}$As- and $^{51}$V-NMR under elevated pressures}


\author{Keiji Ueshima$^1$}
\author{Fei Han$^{2,3}$}
\author{Xiyu Zhu$^{2,3}$}
\author{Hai-Hu Wen$^{2,3}$}
\author{Shinji Kawasaki$^1$}
\author{Guo-qing Zheng$^{1,2}$}

\affiliation{$^1$Department of Physics, and Research Center of New Functional Materials for Energy Production, Storage and Transport, Okayama University, Okayama 700-8530, Japan}
\affiliation{$^2$Beijing National Laboratory for Condensed Matter Physics,
Institute of Physics, Chinese Academy of Sciences, Beijing 100190, China}
\affiliation{$^3$Center for Superconducting Physics and Materials, National Laboratory of Solid State Microstructures and Department of Physics, Nanjing University, Nanjing 210093, China}


\date{\today}

\begin{abstract}
We report $^{75}$As- and $^{51}$V-nuclear magnetic resonance (NMR) measurements on the iron-based superconductor Sr$_2$VFeAsO$_3$  with alternating stacks structure. We find that the $^{75}$As nuclear spin-spin relaxation rate ($1/T_2$) shows a pronounced peak at  $T_N$ = 165 K, below which the resonance peak shifts to a higher frequency due to the onset of an internal magnetic field. The  $^{51}$V spectrum does not shift, but is broadened below $T_N$.
We conclude that the Fe electrons oder antiferromagnetically below $T_N$ with a magnetic moment $m_{Fe}$ $\sim$ 0.4 $\mu_B$. Application of  external pressure up to 2.4 GPa reduces $T_N$ in a rate of $-$40 K/GPa, and enhances the superconducting transition temperature $T_c$ in a rate of  2 K/GPa. The pressure-temperature phase diagram  for Sr$_2$VFeAsO$_3$ shows that superconductivity coexists with antiferromagnetism over a wide pressure range with an unprecedented high $T_c$ up to 36.5 K.
\end{abstract}


\maketitle


\section{Introduction}
The discovery of superconductivity in iron-pnictide LaFeAsO$_{1-x}$F$_{x}$ at the superconducting transition temperature $T_{c}$ = 26 K\cite{Kamihara} has generated much interest, and led to the discovery of many other families.\cite{Chen,Ren,RotterBaK,SefatCo,Wang,Tapp,Hsu,Guo} 
 The parent compounds are metallic and show a tetragonal to orthorhombic structural phase transition followed by an antiferromagnetic order. Superconductivity appears after suppressing these orders by electron/hole doping to the FeAs layer, and/or by applying pressure.\cite{Park,Alireza,Torikachvili} Experiments have suggested that the superconductivity may be related to the quantum critical fluctuations   associated with the above-mentioned orders.\cite{Zhou,Ning,Oka,Nakai,Li3} 
 
Recently, high-$T_c$ superconductivity was found in  Sr$_2$VFeAsO$_3$ that composes of alternating stacks of FeAs and perovskite Sr$_2$VO$_3$ layers.\cite{Wen}
The $T_c^{\rho}$ = 37 K was inferred from the electrical  resistivity measurement, and  $T_c^{\chi}$ = 32 K from the magnetic susceptibility. Hereafter in this paper, the $T_c$ refers to $T_c^{\chi}$.
 It was reported that $T_c$ increases to 43 K at a pressure $P$ = 4 GPa.\cite{Kotegawa} As in the cuprate high-$T_c$ superconductors, this opens a new gateway to studying the relationship between crystal structure and superconductivity, and has triggered the synthesis of other compounds with similar stacking structures.\cite{Ogino,Kakiya}  It is interesting that the superconductivity in Sr$_2$VFeAsO$_3$ is realized without doping by element substitution.  Since most stoichiometric Fe-based compounds host magnetism, searching for a possible magnetic order in Sr$_2$VFeAsO$_3$ has also become an important subject.\cite{Cao}

Indeed, previous magnetic susceptibility and specific heat found an anomaly at a high temperature $T_M$ $\sim$ 150 K, which was ascribed to a possible magnetic transition.\cite{Cao} Since the Fe M\"{o}ssbauer spectra showed no additional broadening below $T_M$, it was proposed that the anomaly originates from the V spin ordering.\cite{Cao}  However, the reported Fe M\"{o}ssbauer spectra split into two peaks even above $T_M$,\cite{Cao} and the reported full width at the half maximum (FWHM) of the M\"{o}ssbauer spectra is rather comparable with that observed in the Fe ordered state of non-doped BaFe$_2$As$_2$  \cite{RotterBa122} and LaFeAsO.\cite{Klauss} This suggests that it is difficult to rule out Fe spin order by the M\"{o}ssbauer measurement, and the origin of the anomalies at $T_M$ $\sim$ 150 K in Sr$_2$VFeAsO$_3$ remains unclear. It is important to elucidate whether the anomaly is due to a magnetic order, and whether the Fe electrons or V electrons order. This is because the relationship between magnetism and superconductivity is an important issue generally in strongly-correlated electron systems.

In this paper, we report $^{75}$As- and $^{51}$V-NMR studies on polycrystalline Sr$_2$VFeAsO$_3$ under elevated pressures. 
Since Sr$_2$VFeAsO$_3$ is a stoichiometric compound, we use external pressure to tune the electronic state. We have constructed a pressure-temperature phase diagram that shed lights on the relationship between magnetism and superconductivity in this system.
  
We found  an internal magnetic field $H_{int}$ = 1.1 T at the As site, while the $^{51}$V-NMR spectrum is broadened due to a field distribution of   0.005 T  below $T_N$ = 165 K. The $^{75}$As spin-spin relaxation rate ($1/T_2$) shows a pronounced peak at $T_N$. We conclude that it is the Fe electrons that order below $T_N$, which produces a large internal magnetic field at the As site and a  field distribution at the V site, since V is far away from Fe in the unit cell.
In the superconducting state, the spin-lattice relaxation rate 1/$T_1$ decreases rapidly below $T_c$ as in other pnictides.\cite{Matano,Ooe,Ning,Oka,Nakai,Li3} 
Thus superconductivity coexists with magnetism in Sr$_2$VFeAsO$_3$ as in other systems,\cite{Zhou,Li} but with a  much higher $T_c$ of 36.5 K.

\section{Experimental}

Polycrystalline samples of Sr$_2$VFeAsO$_3$ were synthesized by using a two-step solid-state reaction method.\cite{Wen} First, SrAs powders were obtained by the chemical reaction method with Sr pieces and As grains. Then, they were mixed with V$_2$O$_5$ (3N), SrO (2N), Fe and Sr powders (3N), in the formula Sr$_2$VFeAsO$_3$, ground and pressed into a pellet shape. The weighing, mixing, and pressing process were performed in a glove box with a protective argon atmosphere (the H$_2$O and O$_2$ contents were both below 0.1 ppm). The pellets were sealed in a silica tube with 0.2 bar of Ar gas and followed by a heat treatment at 1150 $^\circ$C for 40 h. Then it was cooled down slowly to room temperature. The transport and magnetic properties were described previously \cite{Wen}. 
A large pellet ($\sim$ 1000 mg) was crushed into coarse powder for NMR measurements.

The ac susceptibility measurements using the $in$-$situ$ NMR coil at zero magnetic filed indicates $T_c$ = 32 K. $^{75}$As and $^{51}$V-NMR measurements were carried out by using a phase-coherent spectrometer. The NMR spectrum is obtained by integrating the spin echo intensity by changing the resonance frequency ($f$) at a fixed magnetic field of 12.951 T. The  $T_1$ was measured by using a single saturating pulse, and is determined by fitting the recovery curve of the nuclear magnetization to the theoretical function ; ($M_0-M(t)$)/$M_0$ = 0.9$\exp$($-t/T_1$) + 0.1$\exp$($-6t/T_1$), where $M_0$ and $M(t)$ are the nuclear magnetization in the thermal equilibrium and at a time $t$ after the saturating pulse. The spin-spin relaxation rate (1/$T_2$) was obtained from the relation of the spin echo intensity $I(\tau)$ = $I$(0)$\exp$($-2\tau/T_2$) where $\tau$ is the time separation between the $\pi/2$ and $\pi$ pulses. Both $T_1$ and $T_2$ were measured at the spectra peak corresponding to  $H//ab$ configuration.

The pressure was applied by utilizing a NiCrAl/BeCu piston-cylinder type cell filled with silicon oil\cite{PES1} as the pressure-transmitting medium. The pressure at low temperatures was determined from the $T_c$ values of Sn metal measured by a conventional four-terminal method. The pressure distribution $\Delta$$P$ at low temperatures due to solidification of pressure transmitting media for present cell is quantitatively determined to be 5$\%$.\cite{deltaP}

\section{Results}
\subsection{Magnetic order probed by $^{75}$As and $^{51}$V-NMR}
\begin{figure}[h]
\begin{center}
\includegraphics[width=8cm]{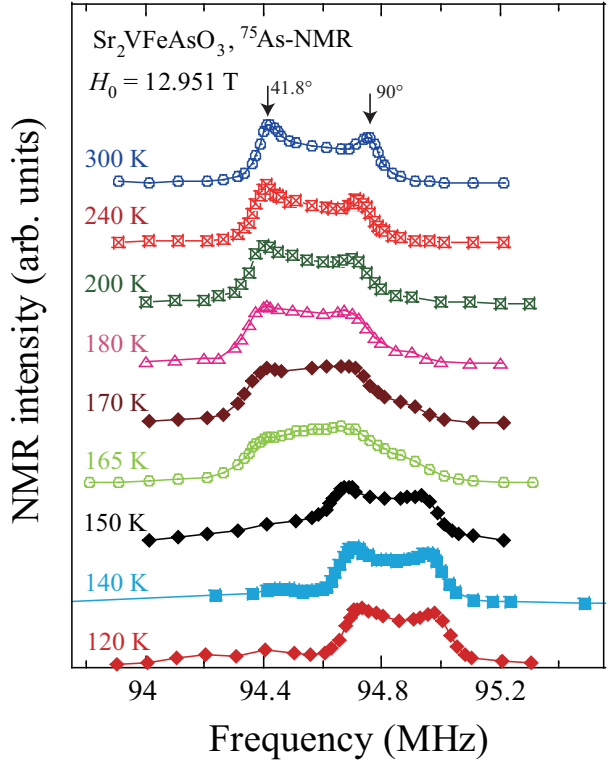}
\caption{(color online) Temperature dependence of the frequency swept $^{75}$As-NMR spectra (center transitions only) at the fixed magnetic field of $H_0$ = 12.951 T. The two horns correspond to $\theta$ = 41.8$^\circ$  and 90$^\circ$ ($H_0 \parallel ab$-plane), see text for detail. }
\end{center}
\end{figure}

First, we show evidence for Fe magnetic order from $^{75}$As-NMR and $^{51}$V-NMR. Figure 1 shows the typical $^{75}$As NMR spectra for the central transition ($m$ = 1/2 $\leftrightarrow$ -1/2 transition). 
The total nuclear spin Hamiltonian is written as $\mathcal{H}$ = $\mathcal{H}_{z}$ + $\mathcal{H}_{Q}$, where $\mathcal{H}_{Z}$ = $\gamma$$\hbar$$\overrightarrow{I}\cdot(\overrightarrow{\textit{H}_{0}}$ + $\overrightarrow{\textit{H}}_{int})$,  
and $\mathcal{H}_{Q}$ is due to the interaction between the nuclear quadrupole moment and the electric field gradient (EFG). $H_{int}$ is a local magnetic field due to the hyperfine interaction, and $H_0$ is the applied magnetic field. 
The sizable $\mathcal{H}_{Q}$ for  $^{75}$As whose nuclear spin  is $I$=3/2 has a large perturbation effect on the $m$ = 1/2 $\leftrightarrow$ -1/2 transition\cite{Abragam}. Since the crystalline axis in a powdered sample is randomly distributed, the NMR spectrum shows a powder pattern depending on the angle $\theta$ between the principal axes of the EFG ($z'$) and the external magnetic filed ($z$).\cite{Abragam} For the case of the As site in the FeAs layer, $z'$ corresponds to the crystal $c$ direction. 

As seen in Fig. 1, there is a two-horns shape  in the frequency dependence of the spectrum. These two horns (peaks) correspond to $\theta$ = 41.8$^\circ$ (peak at lower frequency) and  $\theta$ = 90$^\circ$ i.e. $H_0$//$ab$ (peak at higher frequency). We are the first to observe the NMR spectra around $T$ $\sim$ 150 K, while previous As-NMR attempt failed to detect a signal around this temperature.\cite{Kotegawa}  We find that the spectrum shifts to a higher frequency below $T_N$ = 165 K. The frequency shift of the $\theta$ = 90$^\circ$ peak indicates the appearance of an internal field. 

In contrast, the $^{51}$V-NMR spectrum does not shift, while it is broadened  below $T_N$ = 165 K, 
as seen in  Fig. 2(a).   As seen in the figure, we observed a single peak. Since the nuclear spin for V nucleus is $I$ = 7/2, this suggests that the EFG is small, if any, at the V site.
The spectrum can be fitted by a single Lorentzian curve. As plotted in Fig. 2(b), the FWHM of the $^{51}$V-NMR spectrum is increased below $T_N$ = 165 K. At $T$=100 K, the broadening is equivalent  to  a field distribution of $\sim$ 0.005 T. 

\begin{figure}[h]
\begin{center}
\includegraphics[width=9cm]{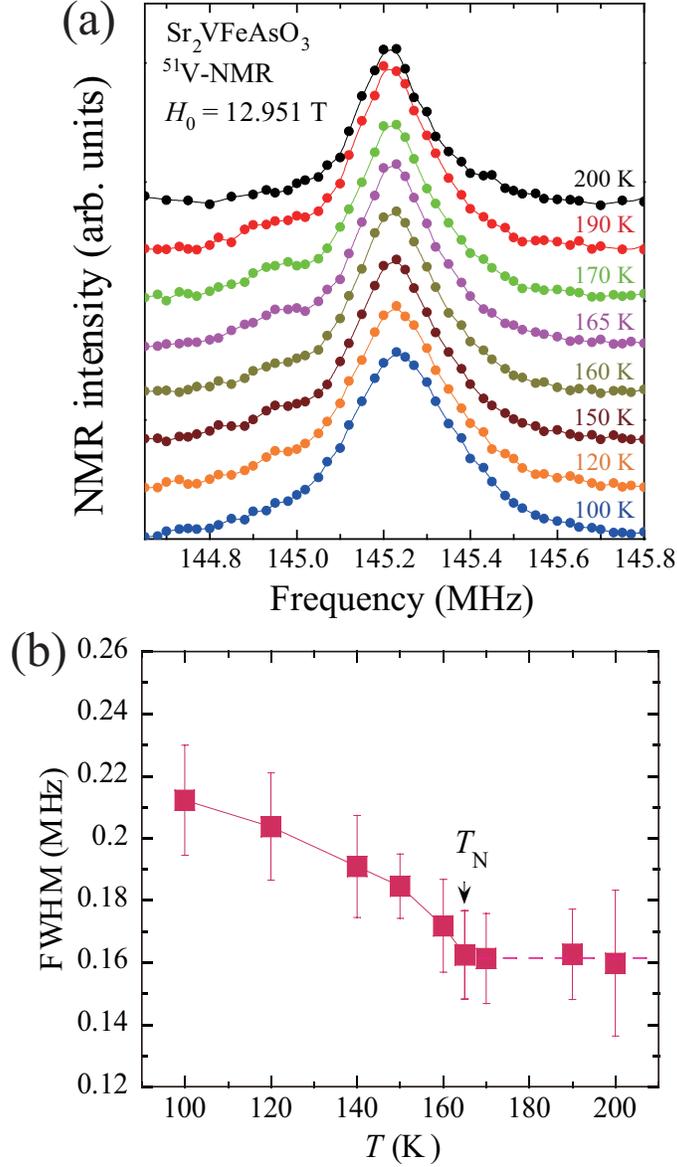}
\caption{(color online) (a) Temperature dependence of the frequency swept $^{51}$V-NMR spectra at the fixed magnetic field of 12.951 T. (b) Temperature dependence of the full width at the half maximum (FWHM).}

\end{center}
\end{figure}

The appearance of the internal magnetic field at the As site and a magnetic field distribution at the V site is due to a  magnetic order, which  is  corroborated by the spin dynamics. Figure 3(a) shows the temperature dependence of $1/T_2$, which exhibits a pronounced peak at $T_N$. Such a peak in a temperature dependence of $1/T_2$ is a characteristic feature of a magnetic order,\cite{Yoshida} since $1/T_2$ probes the longitudinal fluctuating field. At the V site, however, $1/T_2$ shows no anomaly at $T_N$.

The results are consistent with the Fe electrons ordered magnetically below $T_N$. 
Namely, the ordered Fe electrons produce an internal magnetic field at the As site through the Fe-3$d$ and As-2$p$ orbital mixing, and create  a field distribution at the V site since V is far away from Fe in the unit cell.  
Assuming the same Fe spin alignment as in LaFeAsO,\cite{Klauss} the effective field at the As site when the external field is applied along the $a$-axis  is given by  $H_{eff}$  = $\sqrt{H_0^2+H_{int}^2}$, where $H_{int}$ is the internal field produced by the ordered Fe spins.\cite{KitagawaBa,Li} The calculated  $H_{int}(T)$ and the estimated $m_{Fe}$ using the hyperfine coupling constant 
$^{75}A_{hf}^{\perp}$ = 2.64 T/$\mu_{B}$\cite{KitagawaBa} are plotted in Fig. 3(b). The saturated values $m$ $\sim$ 0.4 $\mu_{B}$ is comparable to that observed in non-doped BaFe$_2$As$_2$\cite{Bao} and LaFeAsO.\cite{Dai} Furthermore,  a field distribution at the V site due to a dipole-dipole interaction from the Fe spins ($s_j$),  
$\Delta H_d$=$\displaystyle \sum_{j}\ \{\frac{3(s_j{\cdot}r_j)}{r^5_j}-\frac{s_j}{r^3_j}\}$, is calculated to be $\sim$0.004 T at the V site, which is in reasonable agreement with the observed $\Delta H$. Thus, we conclude that the Fe electrons order  antiferromagnetically  in Sr$_2$VFeAsO$_3$, as in other iron pnictides.

The transition temperature $T_N$ = 165 K found in the present sample is slightly higher than $T_M \sim$ 150 K seen in the susceptibility and specific measurements.\cite{Cao} This is probably due to a difference in the sample quality. In fact, inspection of the derivative of the resistivity data for our sample shows an anomaly at $T_N \sim$ 165 K, being consistent with the NMR results.
The sample dependence of a putative magnetic transition was reported previously by various measurements.\cite{Munevar,Johrendt} For example,  $\mu$SR(Ref.\cite{Munevar}) and neutron scattering measurements (Ref.  \cite{Johrendt}) on a sample with lower $T_c \sim$ 25 K found an anomaly at a rather lower $T_M \sim$ 40 K.
The sample difference is likely due to the deficiency of oxygen. When oxygen is deficient, $T_c$ is reduced.\cite{Wen2} 

\begin{figure}[h]
\begin{center}
\includegraphics[width=8cm]{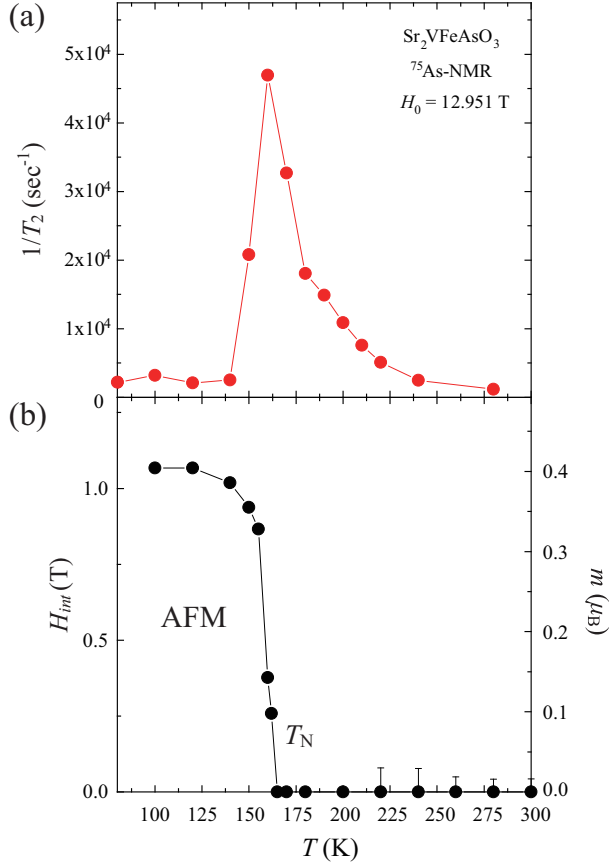}
\caption{(color online)  (a) Temperature dependence of $^{75}$As spin-spin relaxation rate. (b) Temperature dependence of the internal magnetic field ($H_{int}$) (left vertical axis) deduced from frequency shift of NMR spectra in Fig. 1, and the estimated Fe moment (right vertical axis). }

\end{center}
\end{figure}

\subsection{Temperature - Pressure phase diagram}

 Next we present the magnetic and superconducting properties in Sr$_2$VFeAsO$_3$ under high pressures.  Figure 4 shows the temperature dependence of the ac-susceptibility under pressure measured using the $in$-$situ$ NMR coil. The superconducting transitions are clearly seen as rapid decreases of the susceptibility. The $T_c(P)$ are 32 K, 36 K, and 36.5 K at $P$ = 0, 1.9 and 2.4 GPa, respectively. The $T_c$ increases in a rate of 2 K/GPa, which is consistent with previous resistivity measurement under pressure.\cite{Kotegawa}
  
 \begin{figure}[h]
 \begin{center}
 \includegraphics[width=8cm]{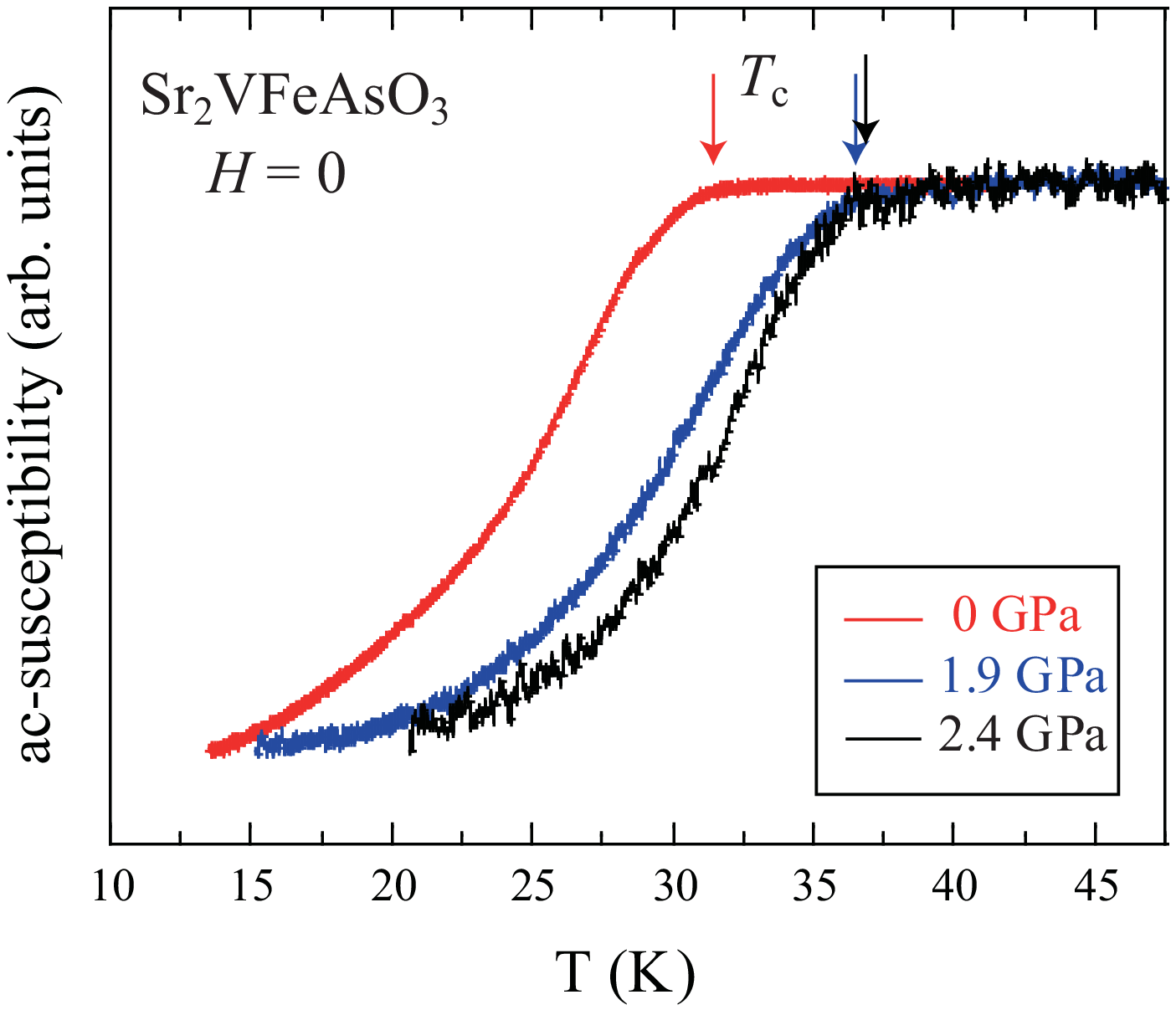}
 \caption{(color online) The ac-susceptibility for  Sr$_2$VFeAsO$_3$ at $P$ = 0, 1.9, and 2.4 GPa, respectively.}
 \end{center}
 \end{figure}

\begin{figure}[h]
\begin{center}
\includegraphics[width=8cm]{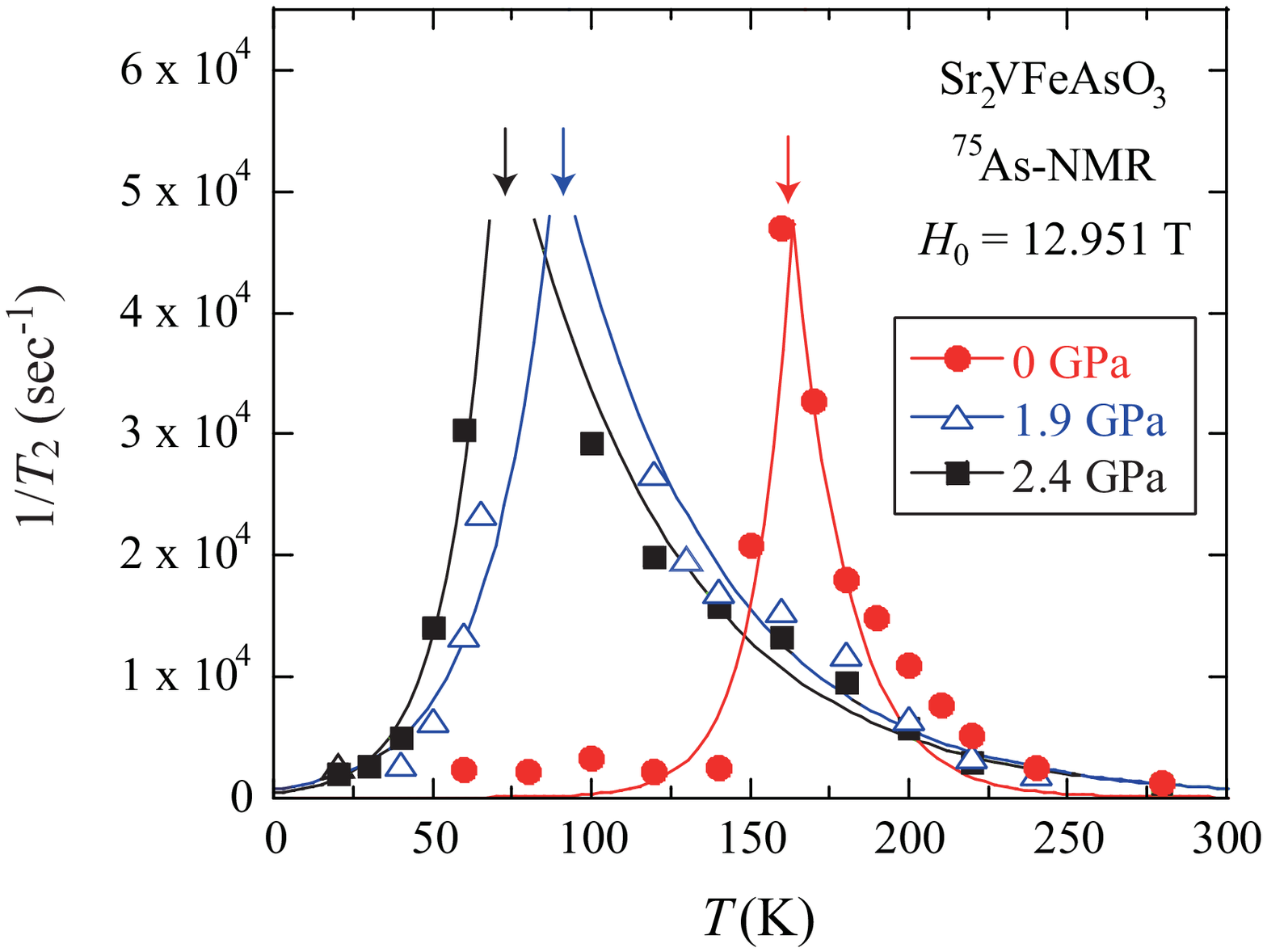}
\caption{(color online) The temperature dependence of 1/$T_2$ at $P$ = 0, 1.9, and 2.4 GPa, respectively. The curves are guides to the eyes. The arrows indicate $T_N$.}  
\label{$T_1$}
\end{center}
\end{figure}

 Figure 5 shows the temperature dependence of $1/T_2$ at $P$ = 0, 1.9, and 2.4 GPa, respectively. Since the sample space in the pressure cell is limited, we can only use a very small amount of sample and as a result, the NMR signal to noise ratio is much smaller compared to the measurement at ambient pressure. It becomes difficult to measure the whole spectra at each temperature around $T_N$. Therefore we adapt the $T_2$ measurement to probe the magnetic transition at high pressures. Even so, the fast spin - spin relaxation around $T_N$ results in a rapid decay of the spin echo, and thus a poor S/N. At ambient pressure, we managed to observe the signal around $T_N$, but this becomes impossible at high pressure because of the small sample amount. Nonetheless, a clear shift of the  data points towards lower temperature can be seen. We estimated $T_N$ by extrapolating the data above and below $T_N (P)$, and obtained  $T_N(P)$ = 90 K $\pm$ 15 K (1.9 GPa) and 70 K $\pm$ 10 K (2.4 GPa)   from the intersection of the extrapolated curves. The error bar corresponds to the half of the temperature interval between which we lost the signal. The $T_N$ is reduced by pressure in a rate of $-$40 K/GPa, which is larger than the cases of SrFe$_2$As$_2$ ($-$25 K/GPa)\cite{KitagawaSr} and LaFeAsO$_{0.945}$F$_{0.055}$ ($-$17 K/GPa) \cite{Khasanov}.

  As seen in Fig. 6, in the superconducting state below $T_c$, $1/T_1$ decreases without a Hebel-Slichter peak as in other iron pnictide superconductors.\cite{Matano,Ooe,Ning,Oka,Nakai,Li3} 
 Since $T_1$ was measured at the peak corresponding to $\theta$ = 90$^\circ$ which experiences an internal magnetic field, this result indicates that superconductivity coexists microscopically with antiferromagnetism. Previous NMR measurements on electron-doped  Ba(Fe$_{1-x}$Ni$_x$)$_2$As$_2$ \cite{Zhou} and  hole-doped Ba$_{0.77}$K$_{0.23}$Fe$_2$As$_2$\cite{Li}  found clear evidence for a microscopic coexistence of antiferromagnetism and superconductivity. However, in those compounds, the coexisting region is rather narrow and $T_c$ there is lower than 20 K . 
 In the present case,  superconductivity is much more robust in the sense that the coexistence takes place in a wider parameter range and  $T_c$ is much higher (as high as 36.5 K) within the antiferromagnetically ordered phase. 

\begin{figure}[h]
\begin{center}
\includegraphics[width=6cm]{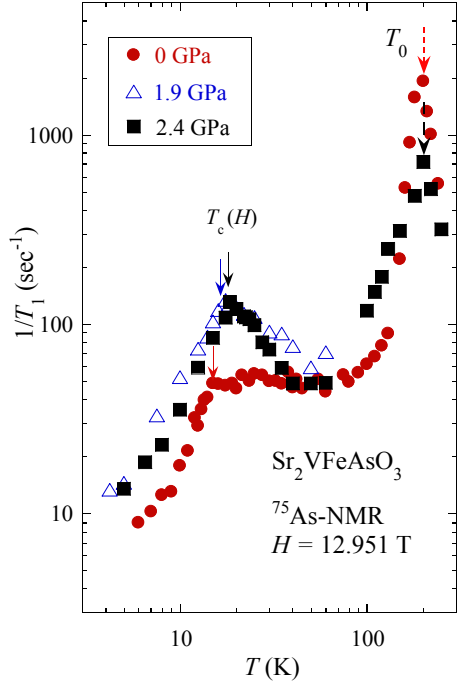}
\caption{(color online) The temperature dependence of $1/T_1$ at $P$ = 0, 1.9, and 2.4 GPa obtained by As-NMR in the fixed magnetic field of 12.951 T. Solid arrows indicate $T_c$($H$).   }
\end{center}
\end{figure}

\section{Discussion}

 The electrical resistivity decreases with decreasing temperature  and shows a further a sharper decreases at $T_0$ $\sim$ 200 K.\cite{Wen} As seen in Fig. 6, we  found a  peak at this $T_0$ in the $T$-dependence of $1/T_1$.  The magnitude of the $1/T_1$ at $T_0$ is reduced by pressure, but  $T_0$ is almost pressure independent. Since no internal magnetic field was found at this temperature, a magnetic cause may be excluded. The origin of $T_0$ is unknown at the present stage.

\begin{figure}[h]
\begin{center}
\includegraphics[width=8cm]{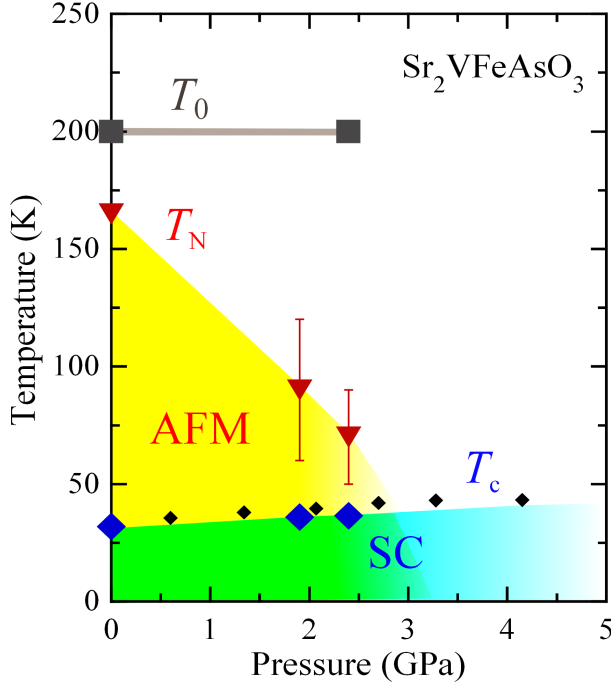}
\caption{(color online) The phase diagram for Sr$_2$VFeAsO$_3$ obtained by present work. The smaller solid diamonds are referred from the resistivity measurements.\cite{Kotegawa}}
\label{Phase diagram}
\end{center}
\end{figure}

Finally, we show in Fig. 7 the pressure-temperature phase diagram for Sr$_2$VFeAsO$_3$ obtained from the present work. The $T_N$ is reduced by pressure rapidly in a rate of $-$40 K/GPa.
Concomitantly, $T_c$ moderately increases in a ratio of 2 K/GPa. Qualitatively, the obtained phase diagram is similar  to other iron pnictides.\cite{SefatCo,Zhou,Li} However, $T_c$ = 36.5 K is much higher than other systems  in the region with a coexistence of antiferromagnetisim.

\section{Summary}
We have performed $^{75}$As- and $^{51}$V-nuclear magnetic resonance (NMR) measurements on the iron-based superconductor Sr$_2$VFeAsO$_3$  with alternating stacks. We found an internal magnetic field at the As site, and a field distribution at the  V site below $T_N$ = 165 K, at which the $^{75}$As nuclear spin-spin relaxation rate ($1/T_2$) shows a pronounced peak. We concluded that Fe electrons order antiferromagnetically with a magnetic moment $m_{Fe}$ $\sim$ 0.4 $\mu_B$. Applying external pressure up to 2.4 GPa reduces $T_N$ in a rate of $-$40 K/GPa, and enhances $T_c$ in a rate of  2 K/GPa. The pressure-temperature phase diagram  for Sr$_2$VFeAsO$_3$ shows that superconductivity coexists with antiferromagnetism with an unprecedented high $T_c$ up to 36.5 K.

\begin{acknowledgments}
We thank S. Tsutsui and P. C. Dai for useful discussions. This work was supported in part by research grants  from MEXT of Japan  (No. 22103004 and 25400374) and from MOST of China (973 project, No. 2011CBA00100 and 2012CB821402).
\end{acknowledgments}


\end{document}